\documentclass[12pt]{article}
\usepackage{amsfonts,amssymb,cite}

\def\noi{\noindent}
\def\Aunames#1{\noi{\bf #1}}
\def\auth#1{${}^{#1}$}
\def\Addresses#1{\medskip\noi \protect
    \begin{description}\itemsep -3pt {\it #1} \end{description}}
\def\addr#1#2{\item[${}^{#1}$]{\it #2}}
\def\email#1#2{\footnotetext[#1]{e-mail: #2}\addtocounter{footnote}{1}}

\newcommand{\bls}[1]{\renewcommand{\baselinestretch}{#1}}

\def\noi{\noindent}
\def\KJ{K_{\rm J}}
\def\RK{R_{\rm K}}
\def\NA{N_{\rm A}}
\def\ur{u_{\rm r}}

\bls{1.01}
\begin{document}
\thispagestyle{empty}

\section*{On new definitions of SI base units.
      Why is the ``atomic'' kilogram preferable}

\Aunames{\small K.A. Bronnikov\auth{a,b,c,1},
             V.D. Ivashchuk\auth{a,b,2},
             M.I. Kalinin\auth{a,3},
             V.V. Khruschov\auth{a,d,4}, \\
             S.A. Kononogov\auth{a,5}, and
             V.N. Melnikov\auth{a,b,6}}

\Addresses{\small
\addr a {Center for Gravitation and Fundam. Metrology, VNIIMS,
    Ozyornaya ul. 46, Moscow 119361, Russia;}
\addr b {Institute of Gravitation and
    Cosmology, PFUR, ul. Miklukho-Maklaya 6, Moscow 117198, Russia;}
\addr c {I. Kant Baltic Federal University, ul. Al. Nevskogo 14,
    Kaliningrad 236041, Russia;}
\addr d {National Research Center Kurchatov Institute,
    pl. Akademika Kurchatova 1, Moscow 123182, Russia}
    }

\begin{abstract}
  We discuss the role of fundamental constants and measurement data for the
  Planck, Avogadro and Boltzmann constants and the elementary electric
  charge in connection with the planned transition to new definitions
  of four base SI units (the kilogram, mole, ampere and kelvin) in terms of
  fixed values of these constants. It is proposed to choose a new
  definition of any base SI unit in terms of a particular fundamental
  physical constant using a number of criteria, or principles, such as
  succession relative to the current SI, a sufficient stability
  of the new unit standards, and concordance between physical dimensions
  of the unit and the corresponding fundamental constant. It is
  argued that a redefinition of the kilogram and mole by fixing the values
  of the atomic mass unit and the Avogadro constant satisfies all these
  criteria and bears some more advantages against the version with
  fixed Planck constant: a well founded approach to definition of the ampere
  and the opportunity to preserve the current relationship between
  definitions of the mole and the kilogram. It is also argued that the
  kelvin can be redefined independently of the other three units.\\

{\bf Keywords:} redefinition of SI base units, fundamental physical
     constant, dimension of a physical quantity, instability of the
     International Prototype of the Kilogram.
\end{abstract}

\email 1 {kb20@yandex.ru}
\email 2 {ivashchuk@mail.ru}
\email 3 {kalinin@vniims.ru}
\email 4 {vkhru@yandex.ru}
\email 5 {kononogov@vniims.ru}
\email 6 {melnikov@vniims.ru}

\section{Introduction}

  There are at present three main unresolved problems in modern
  physics:\\
 --- unification of the four known physical interactions including
  gravity;\\
 --- the present acceleration of the universe, the dark energy (DE) and dark
       ÿmatter (DM) problems;\\
 --- possible variations of fundamental physical constants (FPC): the fine
    structure and gravitational constants, electron to proton mass ratio
    etc.

  The constants we meet in physical theories characterize the stability
  properties of different types of matter, objects, processes, classes of
  processes and so on. These constants are important because they are
  the same under diverse circumstances, at least within the measurement
  uncertainties we have achieved nowadays. That is why they are called FPCs
  and that is why they, or their combinations, can be used as natural scales
  determining the basic units. With scientific progress some theories are
  replaced by more general ones with their own constants, and new relations
  emerge between different constants. Thus what we can discuss is not the
  absolute choice of FPCs but only a choice corresponding to the present
  situation in the physical sciences \cite{1a}.

  At present, the theory of electroweak interactions has a reliable
  confirmation in accelerator experiments with elementary particles. There
  exist a sufficiently well developed and confirmed theory of its
  unification with the strong interaction, the so-called Standard Model, and
  the well-tested (mainly in the Solar System) theory of gravity, Einstein's
  General Relativity (GR). In cosmology, with a rapidly developing
  observational base and its theoretical interpretation, an important role
  belongs to the Standard Cosmological Model (the spatially flat Friedmann
  model with the cosmological constant ($\Lambda$CDM) and its multiple
  modifications and extensions. All these theories, each with its own
  constants, have certain problems \cite{1b}, and a further development of
  physics, especially while solving the above major problems, will probably
  change the presently chosen set of FPCs. When it can happen, we do not
  know.

  The existing systems of units of physical quantities are a part of the
  necessary toolkit of science, and, as any tools, they should adequately
  correspond to the present-day state of the art. Redefinition of the set of
  base units of the International System of Units (SI), being now under
  preparation \cite{1}, can be considered as a response to this challenge.
  The planned revision of SI rests on the proposal to define the base SI
  units by fixing the exact values of the corresponding FPCs, following the
  principle that was already used in the definition of the metre in 1983.

  This revision is under discussion since 2005, and it has been proposed, in
  particular, to fix with zero uncertainty the values of the constants $h,\
  e,\ k$, and $\NA$ and on this basis to redefine the kilogram, ampere,
  kelvin and mole [3--10].
  One of the reasons for changing the existing definitions of these units is
  the revealed temporal instability of the International Prototype of the
  Kilogram (IPK), being on the level of $5\times 10^{-10}$ kg per year
  \cite{8}.

  Employment of exact FPC values has a great significance in metrology
  [12--16], and the proposals to redefine a number of SI
  units on the basis of fixed FPC values have been supported by metrological
  agencies, workshops and conferences \cite{1, 3, 5, 6, 6a}.

  Why was it impossible to introduce such new definitions as soon as their
  necessity was realized? The main obstacle was the insufficient accuracy of
  the current knowledge of the relevant FPCs.

  The current situation with a transition to the new SI (possibly in 2018)
  is reflected in Resolution 1 of the 24th General Conference on
  Weights and Measures (CGPM), ``On possible future revision of the
  International System of Units (SI)'' where it was proposed
  ``{\it
   to continue work towards improved formulations for the definitions of the
   SI base units in terms of fundamental constants, having as far as possible
   a more easily understandable description for users in general, consistent
   with scientific rigour and clarity
   }''.

  The purpose of this paper is to discuss possible modifications of new
  definitions of the four base SI units (the kilogram, mole, ampere and
  kelvin) connected with fixed values of FPCs as well as their realization
  procedure, such as an optimal choice of new definitions of the base
  SI units and their successful employment for the needs of industry,
  commerce and science and for creation of a favourable environment.

  For a new definition of the unit of mass, different versions have been
  proposed; they were considered by international metrological
  organizations and presented at the 23rd and 24th CGPMs. Among the
  criteria for a transition to new definitions of the kilogram and the mole,
  the most important one is achieving the level of $2\times 10^{-8}$
  for the relative standard uncertainty $\ur$ for the values of the Planck
  and Avogadro constants and a consistency between their values obtained by
  different methods with $\ur \simeq 5\times 10^{-8}$. However,
  there are different versions of such new definitions, and even after
  meeting the above requirements, the question of choosing a specific version
  of definitions of the kilogram and mole will remain open. For example,
  for the unit of mass there are versions based on fixing the Planck constant
  and those based on fixing the atomic mass unit. In the present paper, it is
  suggested to choose a new definition of a particular unit of measurement
  with the aid of a number of criteria and principles. When choosing a new
  definition of the mass unit, one should take into account Decision G1
  of the Consultative Committee on Mass and Related Quantities (CCM)
  adopted at a session in 2010 \cite{14}, where it was suggested to soften
  the previously adopted recommendations of BIPM and the 23ed CGPM for
  replacing the IPK \cite{3, 6} as follows:

{\it
--- at least three independent experiments, including work both from watt
    balance and from International Avogadro Coordination projects,
    yield values of the relevant constants with relative standard
    uncertainties not larger than 5 parts in $10^8$. At least one of
    these results should have a relative standard uncertainty not larger
    than 2 parts in $10^8$,

--- for each of the relevant constants, values provided by the different
    experiments be consistent at the 95\% level of confidence,

--- traceability of BIPM prototypes to the international prototype of the
    kilogram be confirmed.
}

  These constraints are the weakest possible since a further increase in
  $\ur$ of measuring the Planck and Avogadro constants would lead to a
  violation of the existing practice of high-precision mass measurements for
  class E1 masses \cite{15}.

  A condition for passing over to a new definition of the kelvin is to
  achieve a level within $\ur(k) \leq 1\times 10^{-6}$ in measurements of
  the Boltzmann constant $k$. Before 2010, the $k$ value, adopted as the
  most accurate one, was obtained in the NIST experiment of 1988, with an
  uncertainty $\ur(k) = 1.7\times 10^{-6}$ \cite{16, 17}. After 2010, the
  effort of a number of research groups from USA, England, Germany, Italy
  and France led to important results in improving the accuracy of
  experimental facilities for measuring the Boltzmann constant.

  A transition to a new definition of the ampere using a fixed value of the
  elementary charge depends, on the one hand, on which decision will be
  made about the new definition of the kilogram (more precisely, whether or
  not the Planck constant value will be fixed); on the other hand, it is
  expected that a quantum standard of the ampere will be created on the
  basis of direct counts of electrons passing a certain surface, [21--31],
  which will allow for introduction of an independent definition of the
  ampere.

  The paper is organized as follows. Section 2 outlines the recent results
  of measurements of the Planck, Avogadro and Boltzmann constants and the
  elementary electric charge. Section 3 suggests and discusses the criteria
  for a preferable choice of new definitions of the base SI units.
  Section 4 discusses the existing problems with a transition to a new
  definition of the kilogram using a fixed value of $h$, known as the
  ``electric kilogram''. In Section 5, on the basis of the criteria
  formulated in Section 3, we suggest to perform a redefinition of the SI
  units with such a set of FPCs that a new definition of the kilogram and
  the mole will employ fixed values of the atomic mass unit (instead of a
  fixed value of the Planck constant) and the Avogadro constant (the version
  known as the ``atomic kilogram).  Section 6 discusses a physical
  foundation of fixing the value of the Boltzmann constant and a possible
  transition to a new definition of the kelvin in the nearest years. Section
  7, the conclusion, summarizes our main proposals. Expansions of abbreviated
  names of institutions mentioned in the text are given before the
  bibliography.

\section {Measurement data}

  At present, three experiments for measuring the Planck and Avogadro
  constants satisfy the first condition of CCM Recommendation G1 of 2010
  \cite{14}. The results of new measurements of the Planck constant at NIST
  with a watt balance (w.b.) were recently published \cite{29},
\[
  h_{\rm NIST{-}14{-}w.b.} = 6.626 069 79(30)
        \times 10^{-34}\ {\rm J\, s},
\]
  with $\ur = 4.5\times 10^{-8}$, and those of the NRC project \cite{30},
\[
     h_{\rm NRC{-}14{-}w.b.} = 6.626 070 34(12)
        \times 10^{-34}\ {\rm J\, s},
\]
  with $\ur  = 2\times 10^{-8}$. These new results in watt-balance
  experiments, as well as the result obtained in the International Avogadro
  project (IAC)in 2011 \cite{31},
\[
     h_{\rm IAC{-}11{-}^{28}Si} = 6.626 070 14(20)
        \times 10^{-34}\ {\rm J\, s},
\]
  also satisfy the second condition of Recommendation G1
  of CCM because the difference $h_{\rm IAC{-}11{-}^{28}Si} - h_{\rm
  NIST{-}14{-}w.b.}$ has turned out to be $5.3\times 10^{-8} h$, while
  $h_{\rm NRC{-}14{-}w.b.} - h_{\rm IAC{-}11{-}^{28}Si}$ is of the order of
  $2\times 10^{-8} h$, and $h_{\rm NRC{-}14{-}w.b.} - h_{\rm
  NIST{-}14{-}w.b.}$ is of the order of $7.3\times 10^{-8} h$. Condition 2
  (G1 CCM) is fulfilled because each of the differences $h(i) - h(j)$ is
  smaller by absolute magnitude than twice the combined standard
  uncertainty,
  $2[(\ur(h(i))h(i))^2 + (\ur(h(j)) h(j))^2]^{1/2}$, $i, j = 1,2,3 (i< j)$.

  The results of the Avogadro project \cite{31} were confirmed with high
  accuracy by new silicon molar mass measurements at NMIJ \cite{N} and
  NIST \cite{V}, for a review see also \cite{Ch}.

  Let us note that CODATA gives for the Planck constant \cite{32} the value
\[
    h_{\rm CODATA{-}2010} = 6.626 069 57(29)\times 10^{-34}\ {\rm J\,s},
\]
  with $\ur = 4.4\times 10^{-8}$, where the old result of NIST (2007) has
   been used.

  When obtaining the Planck constant value in the Avogadro project,
  the value of $\NA$, measured in the same experiment, was used \cite{31}:
\[
    \NA{}_{\rm (Avogadro{-}2010)} = 6.02214084(18)\times 10^{23}\
            {\rm mole}^{-1},
\]
  with $\ur = 3\times 10^{-8}$, and the new value of the Planck
  molar constant \cite{32}
\[
    \NA h_{\rm CODATA{-}2010}= 3.990ÿ312ÿ7176(28) \times 10^{-10}
    {\rm J\,s\,mole^{-1}},
\]
  with $\ur = 0.7\times 10^{-9}$, obtained from the well-known relation
$$
    \NA h = A_{\rm r}(e) M_{\rm u} c \alpha^2/(2 R_\infty), \eqno(2.1)
$$
  where the molar mass constant $M_{\rm u} = \NA m_{\rm u}$ coincides in
  the (``old'') SI with $M_{\rm u0} = 1$ g/mole ($m_{\rm u}
  = m({}^{12}$C)/12 is the atomic mass constant),
  $A_{\rm r}(e)$ is the relative atomic mass of the electron,
  $R_{\infty}$ is the Rydberg constant, $c$ is the speed of light in vacuum,
  and $\alpha$ is the fine structure constant. The calculation of the Planck
  molar constant in CODATA-2010 has used the value of $\alpha$ such that
\[
    \alpha^{-1}{}_{\rm CODATA{-}2010} = 137.035 999 074(44),
\]
  with $\ur(\alpha) = 3.2\times 10^{-10}$ and the electron atomic mass
  value
\[
    A_{\rm r}(e) = 5.485 799 0946\,(22)\times 10^{-4}, \ \
    {\rm with}\ \ \ur (A_{\rm r}(e)) = 4\times 10^{-10}.
\]
  A contribution of the uncertainty of the Rydberg constant,
  $\ur(R_\infty) = 5\times 10^{-12}$ into the relative standard uncertainty
  $\ur(\NA h)$ is negligibly small.

  It should be noted that the quantity $\ur(\NA h)$ also determines a
  constraint on the correction factor $\kappa$, emerging in the new version
  of the definitions of the kilogram and mole based on fixed values of $h$
  and $\NA$:
$$
    \kappa = M_{\rm u}/M_{\rm u0} - 1
      = \NA m(^{12}{\rm C})/(12 M_{\rm u0}) - 1,        \eqno(2.2)
$$
  which is estimated as $|\kappa| < 1.4\times 10^{-9}$ with 95\% confidence
  \cite{32}.

  The emergence of the new parameter $\kappa$ in the new SI is strongly
  criticized in the literature, see \cite{62, 63, 64, 65} and references
  therein. The main arguments come from the chemical and educational
  communities. The introduction of such a new and unnecessary entity is
  evidently out of accord with the famous Occam's razor principle.

  At measurements of the Avogadro constant with the aid of silicon crystal
  balls, the following relation is used \cite{31}:
\[
     \NA = 8M/(\rho a^3),
\]
  where $\rho$, $M$ and $a$ are the density, molar mass and lattice constant
  of silicon, respectively. In this method, the main contributions to the
  full uncertainty budget are connected with measurements of the accuracy
  of the ball's spherical surface and its roughness (65.8\%), the mass
  of the ball's surface layers (16.7\%), the crystal lattice parameter
  (8.7\%) and the molar mass of silicon (4.9\%). Altogether, these
  contributions, among which the largest one is related to the inaccuracy
  of determining the ball's diameter, did not allow for achieving the
  total $\ur$ of $2\times 10^{-8}$, and as a result of completing the second
  stage of Avogadro international project at the end of 2010, the value of
  the Avogadro constant was determined with $\ur$ equal to $3\times 10^{-8}$
  \cite{31}.

  A large role in rising the accuracy of the final result belongs to
  corrections that emerge due to deflections of the basic parameters of
  silicon crystal balls from their assumed perfect behaviour. Such
  deflections include impurities, crystal lattice defects and inhomogeneities
  of the surface layer that affect the measured values of almost all
  parameters. Thus, for instance, the uncertainties emerging while measuring
  the mass of a surface layer of the ball lead to the second largest
  contribution to the full uncertainty after the error in measuring the
  ball diameter; this contribution is close to 17\% \cite{31}. The impurities
  mostly consist of carbon, oxygen and boron. In addition, there appeared
  impurities which had not been anticipated at the beginning of the
  experiment, they emerged at ball surface polishing and consisted of
  metallic contamination of copper and nickel. The metallic contaminations
  very strongly affected the optical characteristics of the surface layer and
  prevented the precision interferometric measurements.

  However, the results obtained in the Avogadro project was verified at PTB,
  where the value of the Avogadro constant was found after removal of copper
  and nickel from the surface of the silicon ball IAC $^{28}$Si AVO28-S8.
  After wet etching and repeated polishing of the ball's surface, its
  diameter decreased by 300 nm and its mass by approximately 9 mg.
  The $\NA$ value measured at PTB coincided with that previously obtained
  in the Avogadro project within $\ur = 3\times 10^{-8}$. Moreover, a
  measurement of the contributions of extraneous chemical elements in the
  silicon specimen used for fabrication of the ball AVO28-S8, conducted at
  INRIM \cite{35}, also confirms the confidence of $\ur = 3\times 10^{-8}$
  in the result of the Avogadro project.

  The measurements performed at NIST in May 2010 have created a shift in the
  $h$ value: $(h/h_{90} - 1)  = 97(37) \times 10^{-9}$ \cite{29}.  A final
  correction of the NIST-14 result has followed from recalibration of the
  platinum-iridium standard K85 at BIPM, which has led to a mass increase of
  K85 by $40 \times 10^{-9}$, and this in turn led to $h/h_{90} - 1 = 137
  \times 10^{-9}$, which is close to the result of \cite{29}, but the authors
  were unable to find a satisfactory explanation of this shift.

  Let us recall that in the case of a radially directed magnetic field
  $B$, for the current value $I$ in a conducting coil with a conductor of
  length $L$, at the ``force'' phase of the experiment, it holds
$
        ILB = mg,
$
  while the ``velocity'' phase of the experiment is described by the relation
$
        U =  BLv,
$
  where $v$ is the velocity in the vertical direction, and $U$ is the
  inductive electromotive force. As a consequence, we obtain the well-known
  relation
$
        IU = mgv,
$
  which for $I = U'/R$ can be rewritten in the form
\[
        mgv = UI = \frac{U_{90}U'_{90}}{R_{90}}
            \frac{K^2_{\rm J-90} R_{\rm K-90}}{4}\,h,
\]
  where $K_{\rm J-90}$ and $R_{\rm J-90}$ are certain conditionally adopted
  constants which in general do not coincide with the Josephson and von
  Klitzing constants, $\KJ = 2e/h$ and $\RK = h/e^2$. It is assumed here
  that $U$ and $U'$ are measured using the voltage standards on the basis
  of the Josephson effect and giving the values $U_{90}$ and $U'_{90}$,
  respectively, while the resistance $R$ is measured by an ohmmeter based
  on the quantum Hall effect, giving as a result $R_{90}$.

  Thus the new result of measuring $h$ published by NIST \cite{29} in 2014
  points at the existence of significant systematic errors in the previous
  NIST result (2007); it also confirms the result of the Avogadro project
  (2011).

  Unlike the NIST-3 experiment described above, the watt balance experiment
  of the Canadian group (NRC) used a permanent magnet. Main contributions to
  the $\ur$ value are here connected with the errors of ``mass exchange'',
  $1.7\times 10^{-8}$, the knife blade hysteresis, $1\times 10^{-8}$,
  the buoyancy force, $0.1\times 10^{-8}$, magnetization, $0.1\times
  10^{-8}$, and ``susceptibility'', $0.1\times 10^{-8}$ \cite{30}.

  It should be noted that the reliability of the NIST-2014 result of
  measuring the Planck constant was verified to a large extent in the
  process of collaboration between NIST and NRC. The result of Canadian
  researchers in $h$ measurement with a watt balance, $h$(NRC-14-w.b.),
  published in 2014, is of great importance since it gives $\ur(h)$ equal to
  $2\times 10^{-8}$ and differs from the result of the Avogadro project by
  the same amount, whereas the difference between $h$(NRC-14-w.b.) and
  $h$(NIST-14-w.b.), equal to $(7.3\times 10^{-8}\,h)$, is smaller that twice
  the combined standard uncertainty equal to $(9.85\times 10^{-8}\,h)$.
  Therefore, the second condition of CCM G1 is fulfilled, which makes it
  possible to begin the redefinition of SI units in the nearest years
  (according to \cite{6a}, in 2018).

  The BIPM recommendations and Resolutions of the 23rd CGPM of 2007 on the
  transition to new definitions of a number of base SI units have promoted
  an increased activity of theoretical and experimental studies in
  thermometry and the related branches of physics and to an improvement of
  methods and means of primary thermometry for measuring the Boltzmann
  constant with accuracies exceeding $10^6$. The experimental groups at
  NPL (England), PTB (Germany), LNE-CNAM (France), NIST (USA), INRIM (Italy)
  have achieved
  a substantial progress in improving the accuracy of measuring the
  Boltzmann constant [47--54].
  The contributions of various
  experimental factors to the uncertainty budget of measuring $k$ were
  thoroughly studied and estimated quantitatively, especially for acoustic
  gas thermometers.  This has led to a substantial decrease in the
  measurement uncertainties when using different types of primary
  thermometers. For acoustic thermometers, $\ur$ was lowered up to $1\times
  10^{-6}$ and less, for other types of thermometers (on the basis of
  measuring the dielectric constant of a gas, the Doppler broadening of gas
  absorption spectra, noise thermometry) up to $\sim 10^{-5}\div 10^{-4}$.

  The authors of \cite{37} (2011) obtained a value of $k$ with the
  uncertainty $\ur(k) = 1.24\times 10^{-6}$. At NPL, using an acoustic
  gas thermometer with a quasispherical cavity, a result was
  obtained in 2013 with $\ur(k) = 0.71\times 10^{-6}$ \cite{45}. At present,
  according to CODATA-2010 \cite{32}, a value of the Boltzmann constant is
  adopted as $k = 1.3806488(13)\times 10^{-23}$ J/K, with the uncertainty
  $\ur(k) = 0.91\times 10^{-6}$, and so the CGPM demands are already fulfilled.

  The introduction of a new definition of the ampere with the aid of a fixed
  value of the elementary charge is also one of the basic tasks of the planned
  reform of the SI system. As is well known, up to now, the currently
  official definition of the ampere in SI is obsolete, resting on the value
  of a force acting between parallel infinite conductors with a current,
  whereas the practice of precision electric measurements employs the
  macroscopic quantum effects, the quantum Hall effect and the Josephson
  effect, characterized by the von Klitzing ($\RK$) and Josephson ($\KJ$)
  constants, respectively:
$$
    \RK = h/e^2,  \qquad \KJ = 2e/h.            \eqno(2.3)
$$
  The most frequently discussed version of the reform contains a proposal
  to fix the values of $h$ and $e$ with zero uncertainty, hence, according
  to (2.3) also the values of the constants $\KJ$ and $\RK$. As a result,
  the units of charge, the coulomb, and the unit of current, the ampere
  equal to a coulomb per second, will also be fixed, thus bringing the
  electric part of the SI system to correspondence with the modern
  practice of measurements.

  However, this scenario faces a number of problems. First, there is some
  reasonable doubt concerning the validity of equations (2.3) for systems
  with a small number of electrons, and for such systems it is necessary
  to have an independent experimental confirmation of these relations.

  Second, such a confirmation is expected in the nearest years on the basis
  of single-electron tunneling \cite{18}; however, this phenomenon will by
  itself lead to an independent version of the definitions of the electric
  charge and current (in other words, to emergence of a quantum standard of
  the ampere), which creates a problem of choosing the formulations of the
  corresponding definitions in the SI system. Indeed, a direct count of
  the number of electrons crossing a certain section of the conductor per
  unit time directly connects the system unit of charge (the coulomb) with
  the elementary charge. The problem of agreement between different versions
  of definitions for the coulomb and the ampere has acquired the conditional
  name of the ``quantum metrological triangle'', whose ``closure'' is one of
  the purposes of the single-electron tunneling experiments. To put it
  simpler, to ``close'' the triangle means to confirm the validity of Ohm's
  law for three quantities: the Josephson voltage, the Hall resistance, and
  the single-electron tunneling current.

  Lastly, third, a new definition of the ampere should become a constituent
  of the unified system of definitions in the new SI that rests on fixed
  values of a certain set of FPCs. However,
  in connection with the problem of choosing a new definition for the
  kilogram (see Sections 4 and 5 of the present paper), in this set of
  constants, the Planck constant $h$ may be lacking. This circumstance even
  more increases the value of a definition of the ampere which will not
  depend on the $h$ value and will be based on precise counts of elementary
  charges crossing a section of a conductor with current per unit time.

  While planning the SI revision, concerning the ampere, the
  task was formulated to obtain $\ur$ of current measurements within
  $10^{-8}$. Studies and experiments aimed at achieving this goal are being
  conducted in many laboratories of the world. Let us mention, in particular,
  such perspective trends as hybrid (metal-semiconductor) electron pumps
  \cite{22}, electron pumps with graphene quantum dots \cite{23}, and the
  European program ``Quantum ampere --- realization of the SI ampere'',
  whose intermediate results have been published in [27--30].
  The available information indicates that, despite significant success of
  the researchers, a new definition of the ampere on the basis of a fixed
  value of the elementary charge cannot be introduced too soon. Thus, one
  can notice that by now measurements of the currents are not considered
  with uncertainties $\ur(J)$ smaller than $10^{-7}$. An accuracy increase
  of one more order of magnitude will naturally require new effort and time.

\section {Criteria for an optimal choice of constants with fixed values
      for redefinitions of measurement units}

  How to choose the set of FPCs and their fixed values in an optimal way
  for creating new definitions of SI units? At first sight it seems that
  a physical constant, whose value is determined by suitable measurements
  for redefinition of a certain SI unit, cannot be fixed because its value
  inevitably contains a measurement uncertainty. However, if the
  uncertainty is smaller than a prescribed limit, which depends on a maximum
  accuracy of measurements conducted with the unit under consideration,
  then for a new definition of this unit it is quite admissible to fix
  the value of the corresponding constant, unless this procedure can
  lead to a contradiction with the usage of other constants with fixed
  values. It is also clear that the status of any FPC is not absolute and can
  change due to generalization of a relevant theory or when the measurement
  accuracy increases.

  Anyway, the key requirement to the new SI is that it should not aggravate
  the situation in any respect for any users as compared with the old SI.
  This means, above all, a necessary succession with respect to the old
  SI, which implies establishing {\it the same set of base and derivative
  units and the same values of all units as they exist by the day of
  revision\/} in order that the whole enormous set of the existing
  measurement data could be preserved without recalculation.

  Second, evidently, the revision of SI should not worsen the stability
  properties of the standards of any units against the old ones. In this
  respect, it is necessary to recall the detected temporal instability of
  the IPK copies for 100 years \cite{8} on the level of $5\times 10^{-10}$
  per year. So the new BIPM prototypes should have a confirmed temporal
  instability not worse than that, and this requirement should hold for any
  new prototype of the kilogram at any time and place. It is clear that
  similar requirements concern all base SI units.

  We here do not discuss FPC variations on the cosmological scale of times and
  distances closely related to the above-mentioned problems of unification of
  interactions, dark energy and dark matter (see, e.g., \cite{32a, 32b, 32c}
  and references therein).
  Such variations, if any, are too small in the present epoch and cannot
  change the stability
  of the basic standards. But they are of great importance for further
  development of fundamental interaction theories \cite{1a, 32b}.

  Third, for establishing a relationship between the system of measurement
  units and physical theories, it is desirable that the number of fixed
  constants were minimum possible, and that the relation between the FPC
  used and the corresponding unit were as simple as possible. The same
  conditions are desirable for successful teaching of the fundamentals
  of metrology at universities and colleges. This goal is achieved if the
  base measurement unit and the corresponding FPC have the same physical
  dimension. That is, if $\rm [D]_{BU}$ is the dimension of the base unit
  and $\rm [D]_{FPC}$ is that of the FPC to be fixed, it is desirable that
$$
    \rm [D]_{BU} = [D]_{FPC}.                          \eqno(3.1)
$$
  However, a literal application of this principle makes it impossible
  to have such base units as the meter and the ampere. Instead, the velocity
  and electric charge units will be base units, contrary to the first
  requirement above and also to the international standard ISO/IEC 80000
  \cite{ISQ} which determines the set of base physical quantities in the
  International System of Quantities (ISQ). To preserve the same set of base
  units of the new SI, the condition (3.1) for choosing the FPC should be
  relaxed \cite{52}. The extended criterion can be formulated as follows:
  {\it a base measurement unit is defined with the aid of an FPC with the
  same dimension or the one that differs from it by a certain power of time:}
$$
    \rm [D]_{BU} = [D]_{FPC}\times [T]^{\it d},           \eqno(3.2)
$$
  {\it where $d$ is a rational number.} Then the unit of length can be, as
  before, defined by fixing the velocity of light, and a new definition of
  the ampere can rest on a fixed value of the elementary charge. Taking into
  account the all-time high accuracy in measuring times or frequencies, the
  possible presence of the additional factor $[{\rm T}]^d$, which is
  necessary for preserving succession with the old SI, will not affect the
  reproduction accuracy of the corresponding unit.

  The maximum possible simplicity of the new SI also assumes that, if
  possible, the base measurement units should be mutually independent, and
  that no new quantities
  (like the correction factor $\kappa$ in Eq.\,(2.2) above) should be
  introduced at the transition to new definitions. It can be said in advance
  that, as follows from the subsequent sections, the ``electric'' kilogram
  violates both these requirements while the ``atomic'' one respects them.

  Thus we suggest the following criteria for choosing the optimal set of
  FPCs to be fixed for using in the new definitions of SI units:

  A. Succession between the old and new definitions,

  B. The stability requirement for transitions of the unit values,

  C. A minimum number of FPCs to be fixed, the absence of new correction
  factors, and

  D. A maximum simplicity of the relationship between a unit and the
  corresponding FPC, hence agreement between their dimensions.

\section {Definitions of the kilogram and mole: the ``electric kilogram''
    and related problems}

  By the time when the ``light meter'' was introduced (1983) and the
  value of the speed of light $c$ was fixed, the conditions had been achieved
  for high-precision measurements of frequencies and the speed of light,
  exceeding by an order of magnitude or more the accuracy of length
  measurements. It was this circumstance that made it possible to fix the $c$
  value. For the four constants proposed to be fixed for redefinition of the
  SI units, viz., $h$, $e$, $k$, and $\NA$, the present state of affairs
  is different, and for each of the constants there is a situation of its
  own. If one simultaneously adopts all new definitions of the SI units by
  fixing $h$, $e$, $k$ and $\NA$, this will lead, above all, to the problem
  of agreement between the values of these constants. Thus, fixing at once
  the values of $h$ and $\NA$, we would obtain a rigid connection between
  the speed of light in vacuum $c$, the molar constant $M_{\rm u}$, the
  relative atomic mass of the electron $A_{\rm r}(e)$, the fine structure
  constant $\alpha$ and the Rydberg constant $R_\infty$ (see the relation
  (2.1)). To avoid such a connection, it is necessary to change the current
  definitions of these constants, e.g., the definition of the molar constant
  $M_{\rm u}$, and this is really suggested when introducing new definitions
  of the kilogram and the mole based on fixed values of $h$ and $\NA$
  \cite{7}. It is certainly a shortcoming of such new definitions because
  it transfers the molar constant $M_{\rm u}$ to the class of variable
  physical quantities ($M_{\rm u} = M_{\rm u0}(1+\kappa)$), where $\kappa$
  is a correction to be determined from experimental data, and its value
  will change as new data emerge. This shortcoming can be avoided by
  introducing a definition of the kilogram on the basis of fixed values of
  the Avogadro constant and the atomic mass unit \cite{52, 53} (see
  Section 5).

  Besides, a definition of the kilogram based on a fixed value of $h$
  (the ``electric kilogram'' \cite{54}) does not satisfy the succession
  criterion A (which would suggest to replace the IPK with a new standard
  having the dimension of mass), and there is no correspondence between the
  dimensions of a measurement unit and the related FPC (criterion D).

  As to the stability requirement (criterion B), one can note that taking
  into account such external factors as microseisms, variable
  electromagnetic fields, the dependence of the gravitational force on the
  place of measurement, etc., on the operation of a watt balance, it is very
  hard to satisfy this requirement at transfers of the value of the new
  kilogram using these complicated electromechanical devices. This inference
  is confirmed by the comparatively large systematic errors which had not
  been taken into account in the NIST experiment of 2007 \cite{29}.

  We have to conclude that the ``electric kilogram'' does not satisfy any of
  the four criteria A--D formulated above for an optimal choice of constants
  with fixed values.

\section {Definitions of the kilogram and mole: the ``atomic kilogram''
    and its advantages}

  Let us consider in more detail the alternative redefinition procedure
  for the units of mass and the amount of substance in terms of fixed
  values of the Avogadro constant $\NA^*$ and the mass of a carbon-12 atom,
  or the atomic mass unit. In doing so, in order to respect the succession
  with the current SI, we suggest to preserve the definition of the molar
  mass of carbon-12, equal to 12 g. We thus satisfy the succession criterion
  and the criterion (3.1) on the dimensions of the measurement unit and the
  corresponding FPC, which role is now performed by the atomic mass unit.
  This procedure simultaneously specifies the microscopic mass standard on
  the basis of the carbon mass and the macroscopic mass standard on the
  basis of two invariants: the mass of a carbon atom and the fixed value of
  $\NA$. Let us note that the mass of a carbon atom is exactly fixed in the
  new mass unit after fixing the Avogadro constant.  We thus suggest that
  the following relation should hold between the fixed numbers $\NA^*$ and
  $N_{\rm kg}^*$:
\[
        \NA^* = 0.012\,N_{\rm kg}^*,
\]
  where the fixed number of carbon atoms $N_{\rm kg}^*$ specifies the
  macroscopic mass unit of the new SI, the kilogram$^*$, and in this case
  the value of the molar constant $M_{\rm u}$ related to the new units,
  kilogram$^*$ and mole$^*$, will remain as it currently is \cite{52, 53}.

  A new definition of the unit of mass can be formulated as follows:

{\it
  The kilogram$^*$, a unit of mass, is the exact mass of $\NA^*/0.012$
  free atoms of carbon-12 at rest, in the ground quantum state.
  }

  This definition corresponds to the following new definition of the mole:

{\it
  The mole$^*$, a unit of the amount of substance, contains $\NA^*$
  structure elements of this substance.
  }

  The suggested definitions of the kilogram and mole agree with the SI
  definitions of these units existing at present and preserve the existing
  connection between them. Consequently, in this case the existing values of
  the molar mass of carbon $M(^{12}{\rm C})$ and the molar mass constant
  $M_{rm u}$, equal to 12 g/mole and 1 g/mole, respectively, are also
  preserved. At the same time, as is easily seen, the above definitions
  of the kilogram and mole can be adopted independently of each other.

  To specify the value of $\NA^*$, one can use some additional conditions.
  Thus, it would be desirable that such widely used mass units of SI and
  CGS, the kilogram and the gram, contain a whole number of carbon-12 atoms.
  In this case the number $\NA^*$ should be a multiple of 12 \cite{52}. And
  certainly, this value should belong to the experimentally determined
  range ($1 \sigma$ range) \cite{31}:
\[
     \NA = (6.02214066 \div 6.02214102) \times 10^{23}.
\]
  We thus obtain the following value of $\NA^*$:
\[
    \NA^* = 602214087869325727188096
\]
  and more concrete definitions of the kilogram and mole:

{\it
  The kilogram is the unit of mass exactly equal to the mass of
  $50184507322443810599008\times 10^3$ free atoms of carbon-12 at rest, in
  the ground quantum state.
  }

{\it
  The mole is a unit of the amount of substance containing exactly
  $602214087869325727188096$ structure elements of this substance.
  }

  These definitions of the kilogram and mole preserve a succession with the
  current SI definitions and do not violate the existing metrological chains
  of transferring the values of the units of mass and amount of substance
  as well as the existing practice of measuring masses, molar masses
  and amounts of substance \cite{8, 15, 32, 55, 56}.

  As follows from the above-said, if this version is used for the definition
  of the kilogram, then a definition of the ampere based on a fixed
  elementary charge value should also become independent of other units
  and rest on the soon expected quantum standard. We believe it is also an
  important advantage: indeed, there is a nonzero probability that the
  metrological triangle will be closed not quite exactly, leading to certain
  corrections to the relations (2.3), which will in turn complicate the task
  of defining the ampere by fixing both $h$ and $e$.

  Thus, in our view, the set of constants to be fixed connected with the
  ``atomic kilogram'' satisfies all the above four criteria A--D and is
  therefore preferable for the planned SI revision.

  It should be noted for completeness that for the version of SI with the
  ``atomic kilogram'' we get for the Planck constant $\ur (h) = \ur(\NA h)$,
  which is $0.7\times 10^{-9}$ by CODATA-10. For the electric and magnetic
  constants $\varepsilon_0$ and $\mu_0$ obeying the relation
  $\varepsilon_0 \mu_0 = 1/c^2$, we get from the definition of the fine
  structure constant, $\alpha = e^2/(2 \varepsilon_0 h c)$, and Eq.\,(2.1)
\[
   \ur (\varepsilon_0) = \ur (\mu_0) = \ur(A_{\rm r}(e)\alpha^3/R_\infty),
\]
  which is about $1 \times 10^{-9}$ according to CODATA-10. In the new SI
  version of \cite{4} we have smaller values $\ur (\varepsilon_0) = \ur
  (\mu_0) = \ur (\alpha) = 3.2 \times 10^{-10}$. But for the accuracies to
  be achieved in the new SI the value  $\ur (\varepsilon_0) = \ur (\mu_0)
  =10^{-9}$ is sufficient.

\section {Redefinition of the kelvin}

  By now, due to efforts of the international metrological organizations
  and national metrological institutes, the first problem of redefinition
  of the unit of thermodynamic temperature, the kelvin, can be considered
  as a solved one. That is, the kelvin should be defined by fixing the exact
  value of the Boltzmann constant.

  This belief is connected with the nature of the Boltzmann constant as a
  conversion factor between two temperature scales [6, 61--63].
  Indeed, the temperature  enters into thermodynamic laws as the
  combination $\theta = kT$ (according to Gibbs \cite{57}, the modulus of
  distribution), and it is this quantity that is measured in experiments.

  It has been shown \cite{60} that in a molecular system of finite size all
  thermodynamic functions are analytic functions of their variables and
  have no discontinuities even in phase transition domains. Mathematically,
  a transition from one phase to another in the space of thermodynamic
  parameters (e.g., temperature and pressure) occurs in a continuous manner,
  though in a very narrow domain of nonzero size. Moreover, the larger is
  the size of a molecular system, the more narrow is this transition
  domain in the thermodynamic parameter space. Only if the size of the
  molecular system tends to infinity, the size of the phase transition
  domain in the space of thermodynamical parameters
  tends to zero in the thermodynamic limit \cite{60}, and
  discontinuities emerge in the equation of state.

  As to real macroscopic systems, it is, strictly speaking, impossible to
  specify for them the temperature of a phase transition or the temperature
  of coexistence of three phases (like the temperature of the triple point
  of water in the energetic scale,
  $\theta_{\rm TPW}$) with zero uncertainty because the transition
  domain in the space of thermodynamical parameters
  has formally a nonzero size. Accordingly, the thermodynamic
  temperature $T_{\rm TPW}$ cannot be specified with zero uncertainty.
  However, the coefficient that connects these two temperature scales can be
  specified exactly. It is this circumstance that underlies the suggested
  redefinition of the thermodynamic temperature unit.

  Though, physically, the thermometric instruments have measurement
  uncertainties greater by many orders of magnitude than the size of the
  above phase transition domain for a real macroscopic system. Therefore, a
  thermometer perceives it as a single point of discontinuity, which is then
  called a phase transition point. One of such points, the triple point of
  water, is presently used for a definition of the kelvin, such that the
  exact value of its thermodynamic temperature is $T_{\rm TPW} = 273.16$ K.

  Measurements of the ``energetic'' temperature of the triple point of
  water $\theta_{\rm TPW}$ by primary thermometers make it possible to find
  experimentally the Boltzmann constant value from the relation $k =
  \theta_{\rm TPW}/273.16$.  The accuracy of such experimental determination
  of $k$ is the same as the accuracy of measuring $\theta_{\rm TPW}$. This
  value of $k$ is after that used in all measurements of the thermodynamic
  temperature, and its uncertainty contributes to the total uncertainty of
  the measured thermodynamic temperature.

  It is clear that fixing the exact value of the Boltzmann constant better
  corresponds physically to the ideas of modern statistical mechanics than
  fixing the exact value of the temperature of a particular phase transition
  in one of the temperature scales \cite{10, 59, 61}.

  The second task for introducing a new definition of the kelvin is finding
  the exact value of the Boltzmann constant. The most direct method of
  determining the $\theta_{TPW}$ value is its theoretical calculation in the
  framework of statistical mechanics \cite{10, 59, 61}. However, by now,
  not only methods of its exact calculation are absent, but there is even no
  method of its approximate calculation with the accuracy achieved in
  measuring $k$ in modern gas thermometers of various types. Therefore, the
  most real way is the experimental determination of the Boltzmann constant
  with the required accuracy. As shown in
  Section 2, this problem is being solved successfully, and it looks quite
  probable that in the nearest three or four years the values of the
  Boltzmann constant will be obtained in a few independent laboratories with
  an accuracy sufficient for making a transition to a new definition of the
  kelvin.

  So there is a substantial difference in the state of affairs with
  redefinition of the kelvin and that with the units of mass and amount of
  substance. Consequently, as soon as the value of
  $k$ is measured with the necessary accuracy, $\ur = 1 \times 10^{-6}$,
  one can introduce a new definition of the kelvin irrespective of whether the
  necessary accuracies of measuring $h$, $\NA$ and $e$ will be achieved by
  that time.

\section {Conclusion}

  We have analyzed the suggested versions of new definitions of the
  kilogram, ampere, kelvin and mole based on fixing the exact values of
  certain FPCs, and formulated some criteria for choosing the set of FPCs
  with fixed values suitable for redefinition of the base SI units. These
  include (A) succession between the new and old definitions, (B) stability
  in transferring the value of a unit, (C) a minimum number of FPCs fixed
  and absence of new correction factors, and (D) agreement in dimensions of
  the base unit and the corresponding FPC. In particular, these criteria are
  satisfied by the mass of carbon-12 and the electron's electric charge. We
  have demonstrated the advantages of a new definition of the kilogram on
  the basis of the atomic mass unit (the ``atomic kilogram''), at which the
  molar mass of the carbon isotope $^{12}$C, equal to 12 g/mole, will be
  preserved.

  Fixing the atomic mass unit and the Avogadro constant $\NA$ would lead to
  quite clear and logically perfect definitions of the kilogram and mole,
  compatible with the definitions of these two units being in force at
  present, and with the current practice of measurements of mass, molar mass
  and amount of substance \cite{52, 62}. Besides, the fixed values of the
  mass of a carbon-12 atom and the Avogadro constant are natural invariants.
  We can note that in 2011 the Commission on Relative Isotope Content and
  Atomic Weights (CIAAW) of the International Union of Pure and Applied
  Chemistry (IUPAC) has supported the definition of the kilogram based on
  the atomic mass unit (dalton) \cite{63}.

  It is known that while the basic physical quantities used in metrology
  are assumed to be independent of each other, for the corresponding
  measurement units such a dependence is still admissible. For instance, in
  the current SI the definition of the mole depends on that of the kilogram,
  and the mole can only be defined after defining the kilogram. However, as
  we saw in Section 5, it is possible to make the units of mass and the
  amount of substance mutually independent and equal in rights but still
  preserve the existing relationship between them.

  The ``electric kilogram'' concept, implying a fixed value of the Planck
  constant $h$, not only violates the four criteria described above,
  but also leads to certain difficulties in defining the ampere.
  The $h$ value being fixed, the ampere is redefined by fixing also
  the electron charge $e$, hence the quantities $2e/h$ and $h/e^2$ will take
  exact values, and the practical electric units $\Omega_{90}$ and $V_{90}$
  will belong to the class of units of the new SI, which is
  favourable for the metrology of electric measurements. However, as argued
  above, all this cannot be realized until the quantum metrological triangle
  is closed, and it can happen that it will not be closed quite exactly.
  Therefore there are serious reasons to attribute the Planck constant to the
  class of electromagnetic quantum constants (as suggested in the ``atomic
  kilogram'' concept), together with the Josephson ($\KJ$) and von Klitzing
  ($\RK$) constants. Then a new definition of the ampere will be provided by a
  new quantum standard on the basis of single-electron tunneling and a fixed
  value of $e$ only and will not depend on the definition of the kilogram and
  the value of $h$. In this case, the values of $\KJ$ and $\RK$ will be
  found by reconciling the results of various experiments with the same or
  better accuracy than the current one. The practice of using fixed values
  of $\KJ$ and $\RK$ can be preserved at a certain accuracy level, leaving
  open the opportunity of measuring $h$ more and more accurately for testing
  the existing theories and a search for new ones.

  Thus the new SI version with the ``atomic kilogram'' makes all the
  four units in question mutually independent.

  Concerning the kelvin, as follows from the discussion in Section 6, it is
  physically more preferable to define it by fixing an exact value of the
  Boltzmann constant as a conversion factor between two temperature scales
  than by fixing the temperature of the triple point of water in one of the
  temperature scales (the thermodynamic one).

  The most accurate values of the Boltzmann constant have been obtained by
  acoustic gas thermometers. Other methods are so far less accurate. All
  experimental results require a thorough analysis
  of the total uncertainty budget, including an accuracy estimate of the
  equation of state of the working substance and confirmation in
  independent experiments using different types of equipment. The current
  dynamics of accuracy increase of the methods and means of primary
  thermometry shows that a reliable achievement of the required measurement
  uncertainty $\ur(k)$ of the Boltzmann constant by the time of the 26th CGPM
  is quite a real task. As soon as this accuracy is achieved, it will be
  possible to introduce a new definition of the kelvin irrespective of
  whether or not there are necessary conditions for introducing new
  definitions of the kilogram, mole and ampere.

  The new SI adoption is now planned to occur at the CGPM in 2018
  \cite{6a}. There is so far enough time to consider the advantages and
  shortcomings of the suggested new definitions of the SI units and their
  different versions (see, e.g., [43--45])
  taking into account the fulfilment of all necessary criteria and further
  progress in reducing the relative uncertainties of the relevant FPC
  measurements, with a hope that there will be the same improvement as at
  the introduction of the new metre in 1983.

\subsection*{Abbreviations}

    BIPM --- Bureau International des Poids et Mesures.\\
    INRiM --- Istituto Nazionale di Ricerca Metrologica, Italy.\\
    LNE-CNAM, France:\\
\phantom{oo} LNE --- Laboratoire National de metrologie et d'Essais, \\
\phantom{oo} CNAM --- Conservatoire national des arts et me'tiers.\\
    NIST --- National Institute of Standards and Technology, USA.\\
    NMIJ --- National Metrology Institute of Japan.\\
    NPL --- National Physical Laboratory, United Kingdom.\\
    NRC --- National Research Council, Canada.\\
    PTB --- Physikalisch-Technische Bundesanstalt, Germany.


\begin{thebibliography}{99}
\bibitem{19}
    H. Scherer and B. Camarota. Quantum metrology triangle experiments:
    a status review. (To be published in Measurement Science and
    Technology Special Issue on ``Electrical quantum standards and
    their role in the SI''). ArXiv: 1204.6500.
\end{thebibliography}
\end{document}